\documentclass[11pt,a4paper]{article}
\usepackage{amsmath,amssymb,url}
\usepackage{graphicx,tabularx}
\usepackage{array,dsfont}
\usepackage{hyperref}
\usepackage{pgf}
\usepackage{pgfkeys}
\usepackage{booktabs}
\usepackage{ifthen}
\usepackage{xspace}
\makeatletter
\newif\if@preliminary
\@preliminaryfalse
\def\preliminary{\@preliminarytrue}
\def\preprintno#1{\def\@preprintno{#1}}
\def\address#1{\def\@address{#1}}
\def\email#1#2{\thanks{\tt #1@{}#2}}
\def\abstract#1{\def\@abstract{#1}}
\renewcommand\abstractname{ABSTRACT}
\newlength\preprintnoskip
\newlength\abstractwidth
\@titlepagetrue
\renewcommand\maketitle{\begin{titlepage}%
  \let\footnotesize\small
  \hfill\parbox{\preprintnoskip}{%
  \begin{flushright}\@preprintno\end{flushright}}\hspace*{1cm}
  \vskip 60\p@
  \begin{center}%
    {\Large\bf\boldmath \@title \par}\vskip 1cm%
    {\sc\@author \par}\vskip 3mm%
    {\@address \par}%
    \if@preliminary
      \vskip 2cm {\large\sf PRELIMINARY DRAFT \par \@date}%
    \fi
  \end{center}\par
  \@thanks
  \vfill
  \begin{center}%
    \parbox{\abstractwidth}{\centerline{\abstractname}%
    \vskip 3mm%
    \@abstract}
  \end{center}
  \end{titlepage}%
  \setcounter{footnote}{0}%
  \let\thanks\relax\let\maketitle\relax
  \gdef\@thanks{}\gdef\@author{}\gdef\@address{}%
  \gdef\@title{}\gdef\@abstract{}\gdef\@preprintno{}
}%
\def\@citex[#1]#2{\if@filesw\immediate\write\@auxout{\string\citation{#2}}\fi
  \def\@citea{}\@cite{\@for\@citeb:=#2\do
    {\@citea\def\@citea{,\penalty\@m}\@ifundefined
       {b@\@citeb}{{\bf ?}\@warning
       {Citation `\@citeb' on page \thepage \space undefined}}%
\hbox{\csname b@\@citeb\endcsname}}}{#1}}
\def\citerange{\@ifnextchar [{\@tempswatrue\@citexr}{\@tempswafalse\@citexr[]}}
\def\@citexr[#1]#2{\if@filesw\immediate\write\@auxout{\string\citation{#2}}\fi
  \def\@citea{}\@cite{\@for\@citeb:=#2\do
    {\@citea\def\@citea{--\penalty\@m}\@ifundefined
       {b@\@citeb}{{\bf ?}\@warning
       {Citation `\@citeb' on page \thepage \space undefined}}%
\hbox{\csname b@\@citeb\endcsname}}}{#1}}
\long\def\@makecaption#1#2{%
  \sbox\@tempboxa{#1: \emph{#2}}%
  \ifdim \wd\@tempboxa >\hsize
    #1: \emph{#2}\par
  \else
    \hbox to\hsize{\hfil\box\@tempboxa\hfil}%
  \fi
  \vskip\belowcaptionskip}
\def\fmslash{\@ifnextchar[{\fmsl@sh}{\fmsl@sh[0mu]}}
\def\fmsl@sh[#1]#2{%
  \mathchoice
    {\@fmsl@sh\displaystyle{#1}{#2}}%
    {\@fmsl@sh\textstyle{#1}{#2}}%
    {\@fmsl@sh\scriptstyle{#1}{#2}}%
    {\@fmsl@sh\scriptscriptstyle{#1}{#2}}}
\def\@fmsl@sh#1#2#3{\m@th\ooalign{$\hfil#1\mkern#2/\hfil$\crcr$#1#3$}}
\makeatother

\newcommand\ltap{\
  \raise.3ex\hbox{$<$\kern-.75em\lower1ex\hbox{$\sim$}}\ }
\newcommand\gtap{\
  \raise.3ex\hbox{$>$\kern-.75em\lower1ex\hbox{$\sim$}}\ }

\newcommand\simge{\mathrel{%
   \rlap{\raise 0.511ex \hbox{$>$}}{\lower 0.511ex \hbox{$\sim$}}}}
\newcommand\simle{\mathrel{
   \rlap{\raise 0.511ex \hbox{$<$}}{\lower 0.511ex \hbox{$\sim$}}}}

\newcommand\be{\begin{equation}}
\newcommand\ee{\end{equation}}
\newcommand\bea{\begin{eqnarray}}
\newcommand\eea{\end{eqnarray}}
\newcommand\ba{\begin{array}}
\newcommand\ea{\end{array}}

\newcommand\whizard{\texttt{WHIZARD}}
\newcommand\ttb{\ensuremath{t\bar{t}}\xspace}
\newcommand\tth{\ensuremath{t\bar{t}H}\xspace}
\newcommand\bbb{\ensuremath{b\bar{b}}\xspace}

\expandafter\def\csname pgf@textdist@protect\endcsname{}
\pgfset{
     number format/.cd,
     min exponent for 1000 sep=4,
     int detect,
  }
\newcommand{\na}{\textcolor{red}{N/A}}
\newcommand{\kfactor}[2]{
  \def\x{#1}
  \def\y{#2}
  \pgfmathparse{(\x)/(\y)}
  \pgfmathprintnumber[fixed,precision=2]{\pgfmathresult}
}
\newcommand{\tableline}[5]{
        \ifthenelse{\equal{#1}{}}
          {\na & \na & \na}
          {\ifthenelse{\equal{#3}{}}
            {
              \ifthenelse{\equal{#5}{}}
              {$#1(#2)$ & \na & \na}
              {$#1(#2) \cdot 10^{-#5}$ & \na & \na}
            }
            {
              \ifthenelse{\equal{#5}{}}
              {$#1(#2)$ & $#3(#4)$ & \kfactor{#3}{#1}}
              {$#1(#2) \cdot 10^{-#5}$ & $#3(#4) \cdot 10^{-#5}$ & \kfactor{#3}{#1}}
            }
          }
}
\newcommand{\tablelinepp}[5]{
        \ifthenelse{\equal{#1}{}}
          {\na & \na & \na}
          {\ifthenelse{\equal{#3}{}}
            {
              \ifthenelse{\equal{#5}{}}
              {$#1(#2)$ & \na & \na}
              {$#1(#2) \cdot 10^{#5}$ & \na & \na}
            }
            {
              \ifthenelse{\equal{#5}{}}
              {$#1(#2)$ & $#3(#4)$ & \kfactor{#3}{#1}}
              {$#1(#2) \cdot 10^{#5}$ & $#3(#4) \cdot 10^{#5}$ & \kfactor{#3}{#1}}
            }
          }
}

\def\bq{\begin{equation}}
\def\eq{\end{equation}}
\def\ba{\begin{eqnarray}}
\def\ea{\end{eqnarray}}

\newcommand{\fb}{{\ensuremath\rm fb}}

\newcommand{\pb}{{\ensuremath\rm pb}}

\begin{document}

\date{\today}

\preprintno{DESY 20-018}
%%% \preprint{hep-ph/xxxxxxx}

\title{Status of the WHIZARD generator for linear colliders}

\author{J\"urgen Reuter
  \footnote{Talk presented at the International Workshop on Future
    Linear Colliders (LCWS2019),Sendai, Japan, 28 October - 1
    November, 2019. C19-10-28.}   
  \email{juergen.reuter}{desy.de}$^a$, 
  Simon Bra{\ss}\email{simon.brass}{desy.de}$^a$,
  Pia Bredt\email{pia.bredt}{desy.de}$^a$,
  Wolfgang Kilian\email{kilian}{physik.uni-siegen.de}$^b$,
  Thorsten Ohl\email{ohl}{physik.uni-wuerzburg.de}$^c$,
  Vincent Rothe\email{vincent.rothe}{desy.de}$^a$,
  Pascal Stienemeier\email{pascal.stienemeier}{desy.de}$^a$}

\address{\it%
$^a$DESY Theory Group, \\
  Notkestr. 85, 22607 Hamburg, Germany
\\[.5\baselineskip]
$^b$University of Siegen, 
 Physics Department, Walter-Flex-Stra{\ss}e 3,
 57068 Siegen, Germany
 \\[.5\baselineskip]
$^c$University of W\"urzburg,
 Faculty of Physics and Astronomy,
 Campus Hubland Nord,
 Emil-Hilb-Weg 22,
 97074 W\"urzburg, Germany
}

\abstract{
  This summarizes the talk given at the LCWS 2019 conference in
  Sendai, Japan, on the progress of the \whizard\ event generator in
  terms of new physics features and technical improvements relevant
  for the physics programme of future lepton and especially linear
  colliders. It takes as a reference the version 2.8.2 released in
  October 2019, and also takes into account the development until
  version 2.8.3 to be released in February 2020.
}

\maketitle

%%%%%%%%%%%%%%%%%%%%%%%%%%%%%%%%%%%%%%%%%%%%%%%%%%%%%%%%%%%%%%%%%%%%%%%%
%%% Text
%%%%%%%%%%%%%%%%%%%%%%%%%%%%%%%%%%%%%%%%%%%%%%%%%%%%%%%%%%%%%%%%%%%%%%%%

\section{Introduction}

\whizard\ is a multi-purpose event generator for collider
physics~\cite{Kilian:2007gr}. It is a very general framework for all
types of colliders, but with a special emphasis on the physics program
at lepton colliders, and has been used for many studies and design
reports for e.g. ILC, CLIC and
FCC-ee~\cite{Fujii:2015jha,deBlas:2018mhx,Baer:2013cma,Behnke:2013lya,Abada:2019zxq}.
Hard scattering process matrix elements are generated with \whizard's
intrinsic (tree-level) matrix element generator
\texttt{O'Mega}~\cite{Moretti:2001zz}, using the color-flow formalism
for QCD~\cite{Kilian:2012pz}. It supports all particles up to spin 2, 
and also fermion-number violating
vertices~\cite{Ohl:2002jp,AguilarSaavedra:2005pw,Hagiwara:2005wg,Kalinowski:2008fk}.
\texttt{O'Mega} can write matrix-element code as compiled process code
(libraries) or as byte-code instructions in the form of a virtual
machine~\cite{Nejad:2014sqa}. The latter produces very small and
efficient matrix element instructions. The NLO automation will be discussed in
Sec.~\ref{sec:nlo}. \whizard\ comes with two different final- and
initial-state parton shower implementations, a $k_T$-ordered shower as
well as an analytic parton shower~\cite{Kilian:2011ka}. For LC
simulations, \whizard\ ships with the final \texttt{Pythia6}
version~\cite{Sjostrand:2006za} for shower and hadronization; it also
has a full-fledged interface to
\texttt{Pythia8}~\cite{Sjostrand:2014zea}. This is very handy as it
directly transfers data between the two event records of the
generators and allows \whizard\ to use all of \texttt{Pythia8}'s
machinery for matching and merching. \whizard\ also automatically
assigns underlying resonances to full off-shell processes and gives
the correct information of resonant shower systems to the parton
shower. 

One of the special features of \whizard\ is its framework for the
support of lepton collider physics, including electron PDFs with
resummation of soft photons to all orders and hard-collinear photons
up to third order in $\alpha$, the generation of ISR photon $p_T$
spectra, sampling of lepton collider beam spectra~\cite{Ohl:1996fi},
proper simulation of polarized beams, crossing angles and
photon-induced background processes. 

\whizard\ has a large number of hard-coded Beyond the Standard Model
(BSM) models. The newest development for new physics, especially
regarding completely general Lorentz tensor structures, will be
described in Sec.~\ref{sec:bsm}.

\section{New physics and technical features}

\subsection{Performance and integration, technical features}

\whizard\ has a very modular infrastructure that allows to easily
exchange different components: there are several different phase-space
algorithms implemented, as well as several different Monte Carlo
integration options. Besides the traditional \texttt{VAMP}
integrator~\cite{Ohl:1998jn}, there is now a conceptually identical
implementation generalized to an MPI-based parallelization. In
contrast to event generation which can always be trivially
parallelized, adaptive phase space integration cannot so easily
parallelized, and is a major bottleneck for high-multiplicity tree-
and especially loop-level processes. This
\texttt{VAMP2} integrator~\cite{Brass:2018xbv} will now be further
improved with a dynamic load balancer that allows for non-blocking
communication between the different workers. The new setup will be
released in version 3.0$\alpha$, cf. below. Even without the load
balancer speed-ups between 10 and 100 are observed, depending on the
complexity of processes. 

Further technical improvements are the finalization of the proper event
headers for the LCIO event interface for the LC software framework, as
well as the completion of the interface to \texttt{HepMC3}. Rescanning
of event files in order to recalculate hard matrix elements without
recalculating the phase space, now also work with beam spectra and
structure functions. Alternative weights (squared matrix elements) can
now be written out not only in LHE and HepMC formats, but also to LCIO.

%%%%%

\subsection{Beyond the standard model physics}
\label{sec:bsm}

Besides of the full SM samples for TESLA, ILC, CLIC and CEPC,
\whizard\ has been extensively used for BSM simulations where it
contains e.g. complete implementations of Little Higgs
models~\cite{Kilian:2003xt,Kilian:2004pp,Kilian:2006eh,Reuter:2012sd,Reuter:2013iya,Dercks:2018hgz}. Another
interesting feature is 
\whizard's ability to calculate unitarity constraints for vector boson
scattering (VBS) and multi-boson processes and to deliver unitarized
\begin{table}  
  %\begin{sidewaystable}[htbp]
\begin{tiny}  
  \def\arraystretch{1.05}
  \begin{tabular}{l l l l}
    \toprule{}%
    Process & $\sigma^{\text{LO}}[\pb]$ & $\sigma^{\text{NLO}}[\pb]$ & $K$ \\ % & References \\
    \midrule{}%   
    $pp\to  jj$ & \tablelinepp{1.157}{2}{1.604}{7}{6} \\ 
    %%% $pp\to  jjj$ & \tablelinepp{8.921}{47}{22.73}{1}{4} \\ 
    \midrule{}%
     $pp\to  Z$  & \tablelinepp{4.2536}{3}{5.4067}{2}{4} \\ 
     $pp\to  Zj$ & \tablelinepp{7.207}{2}{9.720}{17}{3} \\ 
     $pp\to  Zjj$ & \tablelinepp{2.352}{8}{2.735}{9}{3} \\ 
     \midrule{}%
     $pp\to  W^\pm$   & \tablelinepp{1.3750}{5}{1.7696}{9}{5} \\ 
     $pp\to  W^\pm j$ & \tablelinepp{2.043}{1}{2.845}{6}{4} \\ 
     $pp\to  W^\pm jj$ & \tablelinepp{6.798}{7}{7.93}{3}{3} \\ 
     \midrule{}%
     $pp\to  ZZ$ & \tablelinepp{1.094}{2}{1.4192}{32}{1} \\ 
     $pp\to  ZZj$ & \tablelinepp{3.659}{2}{4.820}{11}{0} \\ 
     $pp\to  ZW^\pm$ & \tablelinepp{2.775}{2}{4.488}{4}{1} \\ 
     $pp\to  ZW^\pm j$ & \tablelinepp{1.604}{6}{2.103}{4}{1} \\
     $pp \to W^+W^- (4f)$ & \tablelinepp{0.7349}{7}{1.027}{1}{2} \\
     $pp \to W^+W^-j \;(4f)$ & \tablelinepp{2.868}{1}{3.733}{8}{1} \\
     $pp \to W^+W^+jj$ & \tablelinepp{1.483}{4}{2.238}{6}{-1} \\
     $pp \to W^-W^-jj$ & \tablelinepp{6.755}{4}{9.97}{3}{-1} \\
     \midrule{}%
     $pp \to W^+W^-W^\pm (4f)$ & \tablelinepp{1.309}{1}{2.117}{2}{-1} \\
     $pp \to ZW^+W^-(4f)$ & \tablelinepp{0.966}{2}{1.682}{2}{-1} \\
     $pp \to W^+W^-W^\pm Z (4f)$ & \tablelinepp{0.642}{2}{1.240}{2}{-3} \\
     $pp \to W^\pm ZZZ$ & \tablelinepp{0.588}{2}{1.229}{2}{-5} \\
     \midrule{}%
     $pp\to  \ttb$ & \tablelinepp{4.588}{2}{6.740}{9}{2} \\  
     $pp\to  \ttb j$ & \tablelinepp{3.131}{3}{4.194}{9}{2} \\ 
     $pp\to  \ttb\ttb$ & \tablelinepp{4.511}{2}{9.070}{9}{-3} \\   
     $pp\to  \ttb Z$ & \tablelinepp{5.281}{8}{7.639}{9}{-1} \\
     & \\ & \\ & \\
     \bottomrule{}
  \end{tabular}
  \quad%
 \def\arraystretch{1.05}
  \begin{tabular}{l l l l}
    \toprule{}%
    Process & $\sigma^{\text{LO}}[\fb]$ & $\sigma^{\text{NLO}}[\fb]$ & $K$ \\ % & References \\
    \midrule{}%
    $e^+e^-\to jj$ & \tableline{622.73}{4}{639.41}{9}{} \\ % & \cite{Ellis:1980wv}% \\
     $e^+e^-\to jjj$ & \tableline{342.4}{5}{318.6}{7}{} \\ % & \cite{Bilenky:1994ad}% \\
     $e^+e^-\to jjjj$ & \tableline{105.1}{4}{103.0}{6}{} \\ % & \cite{Signer:1996bf, Signer:1997dm, Nagy:1997mf, Nagy:1997yn%}\\
     $e^+e^-\to jjjjj$ & \tableline{22.80}{2}{24.35}{15}{} \\ % & \cite{Becker:2011vg, Frederix:2010ne%}\\
     %$e^+e^-\to jjjjjj$ & \tableline{3.62}{2}{0.0}{0}{} \\ % &
                                % \cite{Becker:2011vg,
    % Frederix:2010ne%}\\
    \midrule{}%
     $e^+e^-\to\bbb$ & \tableline{92.32}{1}{94.78}{7}{} \\
     $e^+e^-\to\bbb \bbb$ & \tableline{1.64}{2}{3.67}{4}{1} \\
     $e^+e^-\to\ttb$ & \tableline{166.4}{1}{174.53}{6}{} \\ % & \cite{Schmidt:1995mr, Oleari:1997az, Nason:1997nw% \\
     $e^+e^-\to\ttb j$ & \tableline{48.3}{2}{53.25}{6}{} \\ % & \cite{Bernreuther:1997jn, Brandenburg:1997pu, Brandenburg:1999gm% }\\
     $e^+e^-\to\ttb jj$ & \tableline{8.612}{8}{10.46}{6}{} \\  
     $e^+e^-\to\ttb jjj$ & \tableline{1.040}{1}{1.414}{10}{} \\ 
     $e^+e^-\to\ttb \ttb$ & \tableline{6.463}{2}{11.91}{2}{4} \\
     $e^+e^-\to\ttb \ttb j$ & \tableline{2.722}{1}{5.250}{14}{5} \\
     $e^+e^-\to\ttb \bbb$ & \tableline{0.186}{1}{0.293}{2}{} \\
     \midrule{}%
     $e^+e^-\to \ttb H$ & \tableline{2.022}{3}{1.912}{3}{} \\ % & \cite{Dittmaier:1998dz} \\
     $e^+e^-\to \ttb Hj$ & \tableline{0.2540}{9}{0.2664}{5}{} \\
     $e^+e^-\to \ttb Hjj$ & \tableline{2.666}{4}{3.144}{9}{2} \\
     $e^+e^-\to \ttb \gamma$ & \tableline{12.71}{4}{13.78}{4}{} \\
     $e^+e^-\to \ttb Z$ & \tableline{4.64}{1}{4.94}{1}{}\\
     $e^+e^-\to\ttb Z j$ & \tableline{0.610}{4}{0.6927}{14}{}\\
     $e^+e^- \to\ttb Z jj$ & \tableline{6.233}{8}{8.201}{14}{2}\\
     $e^+e^-\to\ttb W^\pm jj$ & \tableline{2.41}{1}{3.695}{9}{4}\\  
     $e^+e^-\to\ttb \gamma \gamma$ & \tableline{0.382}{3}{0.420}{3}{} \\
     $e^+e^-\to \ttb \gamma Z$ & \tableline{0.220}{1}{0.240}{2}{}\\
     $e^+e^-\to\ttb \gamma H$ & \tableline{9.748}{6}{9.58}{7}{2} \\
     $e^+e^-\to \ttb Z Z$ & \tableline{3.756}{4}{4.005}{2}{2} \\
     $e^+e^-\to \ttb W^+ W^-$ & \tableline{0.1370}{4}{0.1538}{4}{} \\
     $e^+e^-\to \ttb H H$ & \tableline{1.367}{1}{1.218}{1}{2} \\
     $e^+e^-\to \tth Z$ & \tableline{3.596}{1}{3.581}{2}{2}\\
    \bottomrule{}
  \end{tabular}
\end{tiny}
\caption{\label{tab:nlo_comp}
  Selection of validated processes at LO and NLO QCD with
  \whizard. $e^+ e^-$ processes (left) are for 1 TeV fixed beams, $pp$
  processes are for 13 TeV. The scale is the scalar transverse energy,
  $H_T$. Jets are clustered with the anti-$k_T$ algorithm and jet
  radius $\Delta R = 0.5$, with cuts of $p_T > 30 GeV$ for the Born
  jets.}
\end{table}
%\end{sidewaystable}
amplitudes for SMEFT dim-6/dim-8 operators and simplified
models~\cite{Beyer:2006hx,Alboteanu:2008my,Kilian:2014zja,Kilian:2015opv,Fleper:2016frz,Brass:2018hfw},
while precision SM predictions for VBS 
can be found in~\cite{Ballestrero:2018anz}. Ongoing work deals with
the automatic calculation of unitarity limits for multiple
(transversal) vector boson production both for hadron and
(high-energy) lepton colliders. 

Nowadays, new physics models are almost exclusively included via
automated interfaces, e.g. to
\texttt{FeynRules}~\cite{Christensen:2010wz,Christensen:2008py}. 
These explicit interfaces have now been superseded by \whizard's
implementation of its UFO~\cite{Degrande:2011ua}
interface. \whizard\ now (with the upcoming 
versions 2.8.3 and 3.0$\alpha$) supports this completely including
spins 1/2, 3/2, 0, 1, 2, 3, 4, 5, automatic construction of 5-, 6-ary
and even higher vertices, fermion-number violating vertices,
four-fermion vertices (and higher), SLHA-type input files for BSM
models and customized propagators defined in the UFO files. This makes
the old interfaces to FeynRules and SARAH~\cite{Staub:2013tta}
deprecated, however, they will be kept for backwards compatibility. 

%%%%%

\subsection{Next-to-leading order QCD automation}
\label{sec:nlo}

\whizard\ started first with hard-coded next-to-leading order (NLO)
projects regarding QED and electroweak corrections for SUSY
production~\cite{Kilian:2006cj,Robens:2008sa} and NLO QCD correction for $pp \to
b\bar{b}b\bar{b}$~\cite{Binoth:2009rv,Greiner:2011mp}. Now,
\whizard\ is based on an automated implementation of the FKS
subtraction algorithm~\cite{Frixione:1995ms}. In this automated
implementation only the virtual amplitudes are external from one-loop
providers (OLP, there are interfaces to
\texttt{Openloops}~\cite{Cascioli:2011va,Buccioni:2019sur},
\texttt{Recola}~\cite{Actis:2016mpe} and
\texttt{GoSam}~\cite{Cullen:2014yla}), while subtraction terms are
automatically generated in \whizard. 
The NLO QCD has been fully validated as can be seen from
Table~\ref{tab:nlo_comp}. First applications of this automated 
interface have been devoted to linear collider top physics in the
continuum~\cite{Nejad:2016bci} and in the threshold
region~\cite{Bach:2017ggt}. These examples also show NLO calculations
with factorized processes as well as NLO QCD decays. Recently, the
selection of heavy-flavor jets in the jet clustering (bottom and charm)
as well as a veto for them has been added, and also the possibility
for photon isolation to separate perturbative QCD from nonperturbative
effects in photon-jet fragmentation. The final validation is being
finished now, there are still a few ongoing issues especially
regarding easier usage, but an alpha version of \whizard\ 3.0
officially releasing NLO QCD automation will be done in March
2020. \whizard\ allows for a completely automatized POWHEG-type
matching (and damping)~\cite{Reuter:2016qbi} to the parton shower (for
final state showering). While the corresponding matching for
initial-state showering is being implemented, the work on NLO
electroweak corrections has been started and first total cross
sections for simple processes are already available. Next steps here
are the complete validation, as well as the proper matching to the
higher-order corrections for incoming electron PDFs. Also, the work
for other NLO matching schemes has started. 

%%%%%%%%%%%%%%%%%%%%%%%%%%%%%%%%%%%%%%%%%%%%%%%%%%%%%%%%%%%%%%%%%%%%

\subsection{Summary and Outlook}

This is a status report of the close-to-final release version
2.8.2/2.8.3 of the \whizard\ version 2 series, showing intense work on
the complete NLO QCD automation, the completion of
automatic generation of arbitrary Lorentz tensor representations and
the UFO interface, and many technical and convenience developments
driven by the upcoming 250 GeV full SM Monte Carlo mass production for
ILC with 2 ab${}^{-1}$ integrated luminosity.

%%%%%%%%%%%%%%%%%%%%%%%%%%%%%%%%%%%%%%%%%%%%%%%%%%%%%%%%%%%%%%%%%%%%%%%%

\section*{Acknowledgments}

This work was funded by the Deutsche Forschungsgemeinschaft under
Germany’s ExcellenceStrategy – EXC 2121 “Quantum Universe” –
390833306. JRR wants to thank the organizers for a fruitful and
interesting conference in Sendai, which was followed by an intense and
very productive LC generator group meeting at University of Tokyo.

%% %%%%%%%%%%%%%%%%%%%%%%%%%%%%%%%%%%%%%%%%%%%%%%%%%%%%%%%%%%%%%%%%%%%%%%%
%% 
%% \appendix
%% 
%% \section{App1}

%%%%%%%%%%%%%%%%%%%%%%%%%%%%%%%%%%%%%%%%%%%%%%%%%%%%%%%%%%%%%%%%%%%%%%%%
%%% References
%%%%%%%%%%%%%%%%%%%%%%%%%%%%%%%%%%%%%%%%%%%%%%%%%%%%%%%%%%%%%%%%%%%%%%%%

\end{document}